\documentclass[12pt]{article}
\usepackage{amssymb}
\usepackage{amsfonts}
\usepackage{amsmath}
\usepackage{latexsym}
\usepackage[T1]{fontenc}
\usepackage[utf8]{inputenc}
\usepackage{lmodern}
\begin{document}
\title{\bf{Critical look at the time--energy uncertainty relations}\footnote{Talk presented at {\em 3rd Jagiellonian Symposium on Fundamental and Applied Subatomic Physics};  June 23 –-- 28, 2019, Krak\'{o}w,   Poland}}
\author{
K. Urbanowski\footnote{e--mail:  K.Urbanowski@if.uz.zgora.pl, $\;$ k.a.urbanowski@gmail.com}\\
University of Zielona G\'{o}ra, Institute of Physics, \\
ul. Prof. Z. Szafrana 4a, 65--516 Zielona G\'{o}ra, Poland.
}


\maketitle

\begin{abstract}
The Heisenberg and Mandelstam--Tamm time--energy uncertainty relations are analyzed.
The conlusion resulting from this analysis
is that within the Quantum Mechanics of Schr\"{o}dinger and von Neumann,
the status of these relations can not be considered as the same as the status
of the position--momentum uncertainty relations, which are rigorous. The conclusion is that
the time--energy uncertainty relations  can not be considered as universally valid.
\end{abstract}


\section{Introduction}

Before the emergence of quantum mechanics, physicists were convinced that always two different physical quantities can  be measured at the same time with any accuracy. Heisenberg analyzing such quantities as a position and a momentum of the moving electron found that   such a belief is wrong  on the quantum level that is in  all cases when a particle  manifests its quantum properties \cite{H,H2}. Results of this Heisenberg's analysis was formulated in the form of mathematical formulae, which are know as the uncertainty relations.
Heisenberg's uncertainty relations describe connections between uncertainties of the position and momentum and also between uncertainties of time and the energy \cite{H}.
These relations
play an important and significant  role in the understanding of the quantum world and in explanations of its properties.
 We have a mathematically rigorous derivation of the position--momentum uncertainty relation but so far within the  Schro\"{o}dinger and von Neumann quantum mechanics
  there does not exist a rigorous derivation of the
time--energy uncertainty relation.
Nonetheless the time--energy uncertainty relation is considered by many authors as having the same status as the position--momentum uncertainty relation and it is often used  as the basis
for drawing far--reaching conclusions regarding the prediction of
the behavior
of some physical systems in certain situations in various areas of physics and astrophysics
and from time to time  such conclusions were considered as the crucial.
So, the time--uncertainty relation still  requires its analysis and checking whether it is correct and well motivated by postulates of quantum mechanics.
We present here an analysis of the Heisenberg and Mandelstam--Tamm time--energy uncertainty relations and show that the time--energy uncertainty relation can not be considered as universally valid.
    In Sections 2 and 3 the reader finds  theory and calculations. Discussion is presented in Sec. 4. Sec. 5 contains conclusions.

\section{Preliminaries: Uncertainty principle}
One of characteristic and the most important consequences of the quantum mechanics is the uncertainty principle.
The most known form of this principle is the Heisenberg uncertainty principle  for the position and momentum \cite{H,H2},
\begin{equation}
\Delta x\, \cdot \, \Delta p_{x}\,\geq \,\frac{\hbar}{2}, \label{H0}
\end{equation}
where $\Delta x$ and $\Delta p_{x}$ are Hesienberg's "{\em uncertainties}". Unfortunately  there is no precise definitions of these "{\em uncertainties}" in \cite{H}. The rigorous definition of uncertainties was proposed  in \cite{Robertson, Schrod-1930}. Following   \cite{Robertson, Schrod-1930} the uncertainty relation
can be written as follows (see e.g. \cite{M}),
\begin{equation}
\Delta_{\phi} x\ \cdot \Delta_{\phi} p_{x}\,\geq \,\frac{\hbar}{2}, \label{H1}
\end{equation}
where $\Delta_{\phi} x$ and $\Delta_{\phi} p_{x}$ are the standard (root--mean--square) deviations: In the general case for an observable $F$ the standard deviation is defined as follows
\begin{equation}
\Delta_{\phi} F = \|\delta F|\phi\rangle\|, \label{DF}
\end{equation}
where
\begin{equation}
\delta F \stackrel{\rm def}{=} F - \langle F\rangle_{\phi}\,\mathbb{I}, \label{dF}
\end{equation}
and $\langle F\rangle_{\phi} \stackrel{\rm def}{=} \langle \phi|F|\phi\rangle$ is the average (or expected) value of an observable $F$ in a system whose state is represented by the normalized vector $|\phi\rangle \in {\cal H}$,
provided that $|\langle\phi|F|\phi \rangle |< \infty$.
Equivalently:  $\Delta_{\phi} F \equiv \sqrt{\langle F^{2}\rangle_{\phi} - \langle F\rangle_{\phi}^{2}}$.
(In Eq. (\ref{H1})  $F$ stands for position  and momentum operators $x$ and $p_{x}$ as well as for their squares). The observable $F$ is represented by a hermitian operator $F$ acting in a Hilbert space ${\cal H}$ of states $|\phi\rangle$. In general, the relation (\ref{H1}) results from basic assumptions of the quantum theory and from the geometry of Hilbert space \cite{Teschl}. Similar relations hold for any two observables, say $A$ and $B$, represented by non--commuting hermitian operators $A$ and $B$ acting in the Hilbert space of states (see \cite{Robertson,Schrod-1930}), such that $[A,B]$ exists and $|\phi\rangle \in {\cal D}(AB) \bigcap {\cal D}(BA)$, (${\cal D}({\cal O})$ denotes the domain of an operator $\cal O$ or of a product of operators):
\begin{equation}
\Delta_{\phi} A \cdot \Delta_{\phi} B\;\geq\;\frac{1}{2} \left|\langle [A,B] \rangle_{\phi} \right|.\label{R1}
\end{equation}
The inequality (\ref{R1}) is rigorous and its derivation simple. Indeed, let us consider two observables represented by noncommuting operators $A$ an $B$.
Then if to apply  the definition (\ref{dF})  to operators $A$ and $B$, respectively,  one finds that
\begin{equation}
[A,B] \equiv [\delta A, \delta B] \neq 0. \label{[A,B]}
\end{equation}
Hence for all $|\phi\rangle \in {\cal D}(AB) \bigcap {\cal D}(BA)$,
\begin{eqnarray}
\left|\langle \phi| [A,B] |\phi \rangle \right|^{2} &\equiv & \left|\langle \phi| [\delta A,\delta B] |\phi \rangle \right|^{2} \nonumber \\
& \equiv & \left| \,\langle \phi| \delta A \;\delta B |\phi \rangle - \langle \phi| \delta B\; \delta A |\phi \rangle \,\right|^{2} \nonumber \\
& =  &  \left|\, \langle \phi| \delta A \;\delta B |\phi \rangle  \, - \, \left( \langle \phi| \delta A\; \delta B |\phi \rangle \right)^{\ast} \right|^{2} \nonumber \\
& =  & 4\,\left|\,{\rm Im.}\left[\langle \phi| \delta A \;\delta B |\phi \rangle \right]\, \right|^{2} \nonumber \\
&\leq& 4 \,\left| \,\langle \phi| \delta A \;\delta B |\phi \rangle \,\right|^{2} \nonumber \\
& \leq & 4 \,\left\| \delta A |\phi \rangle \right \|^{2}\,\cdot\,\left \|\delta B|\phi \rangle \right\|^{2}  \nonumber \\
& \equiv & 4 \;(\Delta_{\phi}A)^{2}\,\cdot \,(\Delta_{\phi}B)^{2}, \label{H-ku}
\end{eqnarray}
which reproduces inequality (\ref{R1}). It is because $ (\Delta_{\phi} A)^{2} \equiv \left\|\delta A|\phi\rangle \right\|^{2}$ and $ (\Delta_{\phi} B)^{2} \equiv \left\|\delta B|\phi\rangle \right\|^{2}$. The above derivation seems to be the simplest one. \\

One can find in the literature and in many textbooks the other derivations of  the Heisenberg inequality. For completeness they are presented below in short.
The most commonly used methods of deriving the uncertainty relation (\ref{R1}), which can be found in the literature, are the following:
The first one follows the Robertson \cite{Robertson} and Schr\"{o}dinger \cite{Schrod-1930}. This methods uses the obvious relation resulting from the Schwartz inequality,
\begin{equation}
\left\| \delta A |\phi\rangle \right\|^{2}\;\left\| \delta B|\phi\rangle \right\|^{2}\,\geq \,\left|\langle\phi| \delta A\;\delta B|\phi \rangle \right|^{2}. \label{dAdB}
\end{equation}
The next step within this method is (see, e.g., \cite{M,Teschl,Merzb,Zet} and so on) to write  the product $\delta A\;\delta B$ as a combination of the  hermitian and anti--hermitian parts,
\begin{equation}
\delta A\;\delta B\,=\, \frac{\delta A\;\delta B\,+\,\delta B\;\delta A}{2} \,+\,i\,\frac{(-i)(\delta A\;\delta B\,-\,\delta B\;\delta A)}{2}. \label{dAdB-1}
\end{equation}
Here
\begin{equation}
\{(-i)(\delta A\;\delta B\,-\,\delta B\;\delta A)\}^{+} \,=\,[(-i)(\delta A\;\delta B\,-\,\delta B\;\delta A)] \,\equiv\,(-i)[A,B] \label{hermit-1}
\end{equation}
is the hermitian operator.
Hence
\begin{equation}
\left| \langle\phi| \delta A\;\delta B| \phi \rangle \right|^{2} = \frac{1}{4}\left|\langle \phi|(\delta A\;\delta B\,+\,\delta B\;\delta A)|\phi\rangle\right|^{2}  + \frac{1}{4} \left|\langle \phi|[A,B]|\phi\rangle \right|^{2}. \label{Re+Im-1}
\end{equation}
Now in order to obtain the desired result one should ignore the first component of the right hand side of the above equation which leads to the following inequality
\begin{equation}
\left|\langle\phi| \delta A\;\delta B|\phi \rangle \right|^{2} \geq \frac{1}{4} \left|\langle \phi|[A,B]|\phi\rangle \right|^{2} \label{U-2}
\end{equation}
and finally
\begin{equation}
\left\| \delta A |\phi\rangle \right\|^{2}\;\left\| \delta B|\phi\rangle \right\|^{2}\,\geq \,\left|\langle\phi| \delta A\;\delta B|\phi \rangle \right|^{2} \geq
\frac{1}{4} \left|\langle \phi|[A,B]|\phi\rangle \right|^{2}, \label{R-Sch}
\end{equation}
which is the proof of the inequality (\ref{R1}).

The similar simpler version of the above proof can be found eg. in \cite{Hall,Grif}. This proof also makes use of the inequality (\ref{dAdB}) but the estimation of the right hand side of this inequality is simpler. Namely within this proof expression $\left|\langle\phi| \delta A\;\delta B|\phi \rangle \right|$ is written as follows
\begin{equation}
\left| \langle\phi| \delta A\;\delta B|\phi \rangle \right|^{2} = \left[ {\rm Re.} ( \langle\phi| \delta A\;\delta B|\phi \rangle )  \right]^{2}\,+\,
\left[ {\rm Im.} ( \langle\phi| \delta A\;\delta B|\phi \rangle )  \right]^{2}. \label{mod=}
\end{equation}
Next  ignoring the  contribution coming from the real part of the above expression one obtains
\begin{equation}
\left| \langle\phi| \delta A\;\delta B|\phi \rangle \right|^{2} \;\geq\;
\left[ {\rm Im.} ( \langle\phi| \delta A\;\delta B|\phi \rangle )  \right]^{2} \label{mod>Im}
\end{equation}
but
\begin{eqnarray}
{\rm Im.} ( \langle\phi| \delta A\;\delta B|\phi \rangle ) &=& \frac{1}{2i} \left(\langle\phi| \delta A\;\delta B|\phi \rangle\, -\, \langle\phi| \delta A\;\delta B|\phi \rangle^{\ast} \right) \nonumber \\
&=&  \frac{1}{2i} \left(\langle\phi| \delta A\;\delta B|\phi \rangle\, -\, \langle\phi| \delta B\;\delta A|\phi \rangle \right) \nonumber \\
&\equiv & \frac{1}{2i} \langle\phi|[ A, B]|\phi \rangle. \label{Im=A,B}
\end{eqnarray}
Thus
\begin{equation}
\left| \langle\phi| \delta A\;\delta B|\phi \rangle \right|^{2} \;\geq\; \frac{1}{4} \left|\langle\phi|[ A, B]|\phi \rangle \right|^{2}. \label{mod>A,B-1}
\end{equation}
This result together with (\ref{dAdB}) means that
\begin{equation}
\left\| \delta A |\phi\rangle \right\|^{2}\;\left\| \delta B|\phi\rangle \right\|^{2} \,\geq \, \frac{1}{4} \left|\langle \phi|[ A, B]| \phi \rangle \right|^{2}. \label{R-u-2}
\end{equation}

The another method to proof the Heisenberg inequality one can find eg. in \cite{Bellac,Galindo}. Within this method using selfadjoint operators $\delta A$ and $\delta B$ one builds a new non--selfadjoint operator
\begin{equation}
L_{\lambda} = \delta A\,+\,i\lambda\,\delta B \neq L_{\lambda}^{+} \label{L-1}
\end{equation}
where $\lambda = \lambda^{\ast} \geq 0$. Then
\begin{eqnarray}
\left\|L_{\lambda}|\phi \rangle \right\|^{2} &=& \left\|\delta A|\phi \rangle\right\|^{2}\,+\,i\lambda\,\left(\langle \phi|(\delta A\,\delta B - \delta B\,\delta A)|\phi \right)\, +\,\lambda^{2}\,\left\|\delta B|\phi \rangle \right\|^{2} \nonumber \\
& \equiv & \lambda^{2}  \left\|\delta B|\phi \rangle\right\|^{2}\,+\,\lambda\,\langle \phi |(+i [A,B])|\phi\rangle\, +\, \left\|\delta A|\phi \rangle\right\|^{2} \geq 0. \label{L-2}
\end{eqnarray}
Note that $(+i [A,B])= (+i [A,B])^{+}$ therefore  the average value of $(+i [A,B])$ is a real number:
$\langle \phi |(i [A,B])|\phi\rangle = (\langle \phi |(i [A,B])|\phi\rangle)^{\ast}$. So we have the second--degree polynomial in $\lambda$, which is positive for any $\lambda$. This implies that discriminant of this equation can not be a positive
\begin{equation}
\left(\langle \phi |(i [A,B])|\phi\rangle \right)^{2}\;-\;4\, \left\|\delta A|\phi \rangle\right\|^{2}\;\left\|\delta B|\phi \rangle\right\|^{2}\,\leq\,0, \label{disc-1}
\end{equation}
which again reproduces the inequality (\ref{R1}).

Summarizing the above part of the analysis
of  the Heisenberg uncertainty relations, let us note that
from (\ref{dAdB}) if follows that the equality in the uncertainty relation (\ref{R1}) takes place when $|\phi \rangle$ is an eigenvector of $A$ or $B$ or when
vectors $\delta A |\phi \rangle$ and $\delta B|\phi \rangle $ are parallel: $\delta A |\phi \rangle \,\parallel \,\delta B|\phi \rangle $ but from (\ref{mod=}) it follows that this equality is possible only if additionally ${\rm Re.}\,( \langle\phi| \delta A\;\delta B|\phi \rangle ) = 0$. From these conditions the following conclusion results:
The necessary nad sufficient condition for the minimum uncertainty, that is for the equality in the uncertainty relation (\ref{R1}) is
\begin{equation}
\delta B |\phi \rangle  \,=\, i\kappa\,\delta A|\phi \rangle, \label{u=}
\end{equation}
where $\kappa \in \mathbb{R}$.  In particular for $A = x$ and $B =p$, that is for the position--momentum uncertainty relation,  solutions of  the criterion (\ref{u=}) are the so--called {\em coherent states}, which have the Gaussian form.

The defect, or perhaps the weakness,  of methods   (\ref{dAdB}) ---  (\ref{disc-1})  of deriving of the uncertainty relation (\ref{R1}) is that using them one should know
in advance what a result  should be  obtained and which components appearing in the intermediate equations during the derivation process should be ignored. For example following the method (\ref{dAdB}) --- (\ref{R-Sch}) in order to obtain the desired result one should know earlier that the hermitian part in (\ref{Re+Im-1}) should be ignored.
Using the approach described by relations (\ref{mod=}) --- (\ref{R-u-2}) one should know in advance that the result (\ref{R1}) can be obtained by ignoring in (\ref{mod=}) the contribution coming from the real part of scalar product $\langle\phi| \delta A\;\delta B|\phi \rangle $. Finally using the method analyzed in (\ref{L-1}) --- (\ref{disc-1}) one should guess the form of  such an auxiliary operator $L_{\lambda}$ (see (\ref{L-1})), and then one should know in advance that  in order to proof the relation (\ref{R1}) the
$\left\| L_{\lambda}|\phi\rangle \right\|^{2}$ should be considered. Comparing  methods described in relations (\ref{dAdB}) ---  (\ref{disc-1})   with the method  (\ref{H-ku}) one can see that the derivation (\ref{H-ku}) of the relation (\ref{R1}) is much shorter and simpler than the others and it does not need any hidden assumption making in advance to find a desired result.\\

\section{Analysis of the Heisenberg and Mandelstam--Tamm time--energy uncertainty relations }

Heisenberg in \cite{H} postulated also the validity of the analogous relation to (1) for the time and energy
(see also \cite{J}).
This relation was a result of his heuristic considerations and it is usually written as follows
\begin{equation}
\Delta_{\phi} t \cdot \Delta_{\phi} E \geq \frac{ \hbar}{2}.\label{H2}
\end{equation}
The more rigorous derivation of this relation was given by Mandelstm and Tamm \cite{M-T} and now it is known as the Mandelstam--Tamm time--energy uncertainty relation. Their derivation is reproduced in \cite{M} and goes as follows: In the general relation (\ref{R1}) the operator $B$ is replaced by the selfadjoint non--depending on time Hamiltonian $H$ of the system considered and $\Delta_{\phi} B$ is replaced by $\Delta_{\phi} H $ and then identifying the standard deviation $\Delta_{\phi} H $ with $\Delta_{\phi} E$ one finds that
\begin{equation}
\Delta_{\phi} A \cdot \Delta_{\phi} E\;\geq\;\frac{1}{2} \left| \langle [A,H] \rangle_{\phi} \right|,\label{M1}
\end{equation}
where it is assumed that $A$ does not depend upon the time $t$ explicitly, $|\phi\rangle \in {\cal D}(HA) \bigcap {\cal D}(AH)$,  and $[A,H]$ exists.
The next step is to use the Heisenberg representation and corresponding equation of motion which allows
to replace the average value of the commutator standing in the right--hand side of the inequality (\ref{M1}) by the derivative with respect to  time $t$ of the expected value of $A$,
\begin{equation}
\langle [A,H] \rangle_{\phi}  \equiv i\hbar  \frac{d}{dt} \langle A \rangle_{\phi}. \label{M2}
\end{equation}
Using this relation one  can  replace the inequality  (\ref{M1}) by the following one,

\begin{equation}
\Delta_{\phi} A \cdot \Delta_{\phi} E\;\geq\;\frac{\hbar }{2} \left|  \frac{d}{dt} \langle A \rangle_{\phi} \right|.\label{M3}
\end{equation}
(Relations (\ref{M1}) --- (\ref{M3}) are rigorous). Next authors \cite{M,M-T} and many others divide both sides of the inequality (\ref{M3}) by the term $\left|  \frac{d}{dt} \langle A \rangle_{\phi} \right|$, which leads to the following relation
\begin{equation}
\frac{\Delta_{\phi} A}{\left|  \frac{d}{dt} \langle A \rangle_{\phi} \right|} \cdot \Delta_{\phi} E \;  \geq \; \frac{\hbar}{2}, \label{M4}
\end{equation}
or, using
\begin{equation}
 \tau_{A} \stackrel{\rm def}{=} \frac{\Delta_{\phi} A}{\left|  \frac{d}{dt} \langle A \rangle_{\phi} \right|},
 \end{equation}
 they come to the final result known as the Mandelstam--Tamm time--energy uncertainty relation,
 \begin{equation}
 \tau_{A} \cdot \Delta_{\phi} E \geq \frac{\hbar}{2}, \label{M5}
 \end{equation}
 where $\tau_{A}$ is usually considered as a time characteristic of the evolution of the statistic distribution of $A$ \cite{M}. The time--energy uncertainty relation (\ref{M5}) and the above described derivation of this relation is accepted by many authors analyzing this problem or applying this relation (see, e.g. \cite{JB,Bauer,Gislason,skr} and many other papers). On the other hand there are some formal controversies
regarding the role and importance of the  parameter $\tau_{A}$ in (\ref{M5}) or $\Delta t$ in (\ref{H2}). These controversies are caused by the fact that in the quantum mechanics the time $t$ is a parameter. Simply it  can not be described by the hermitian operator, say $T$,  acting in the Hilbert space of states (that is time can not be an observable) such that $[H,T] = i\hbar \mathbb{I}$ if the Hamiltonian $H$ is bounded from below.
This observation was formulated by Pauli \cite{Pauli} and it is know as "Pauli's Theorem" (see, eg. \cite{JB,Busch}).  Therefore  the status of the relations (\ref{H2})
and relations (\ref{H1}), (\ref{R1}) is not the same
regarding the basic principles of the quantum theory (see also discussion, e.g.,  in \cite{Vor,Hi1,Hi2,Br}).

The Pauli's conclusion follows from the following analysis:
If  $T = T^{+}$ and $[H,T] = i\hbar \mathbb{I}$ then
\begin{equation}
[H, T^{n}]= i \hbar n \,T^{n-1} \equiv i \hbar \frac{\partial }{\partial T} T^{n}.
\end{equation}
 From this last relation it follows that
\begin{equation}
\left[ H, e^{\textstyle{-i \frac{\lambda}{\hbar} T}} \right]\, =\, \lambda \, e^{\textstyle{-i \frac{\lambda}{\hbar} T}}.
\end{equation}
The consequence of the above commutation relations is that for all $\lambda \in \mathbb{R}$
\begin{equation}
e^{\textstyle{+ i \frac{\lambda}{\hbar} T}}\; H \; e^{\textstyle{-i \frac{\lambda}{\hbar} T}}\;=\;H\,+\,\lambda\mathbb{I}\; \stackrel{\rm def}{=} \;H_{\lambda}. \label{H-lambda}
\end{equation}
From the Stone theorem we know that if $T$ is a selfadjoint operstor,  $T = T^{+}$, then the operator
\begin{equation}
U_{\lambda} \stackrel{\rm def}{=} e^{\textstyle{+ i \frac{\lambda}{\hbar} T}}, \label{U-lambda}
\end{equation}
is the unitary operator: $U_{\lambda}\;U_{\lambda}^{+}\;=\;U_{\lambda}^{+}\;U_{\lambda}\;=\;\mathbb{I}$. So, operators $H$ and $H_{\lambda}$ are unitarily
equivalent  and hence they must have the same spectrum. From (\ref{H-lambda}) it follows that they commute: $[H,H_{\lambda}]=0$ and thus they have common eigenfunctions. The spectrum of $H_{\lambda}$ ranges over the whole real line $\mathbb{R}$ but, by assumption (see, e.g. \cite{Busch}) the spectrum of $H$ is bounded from below. Therefore $H_{\lambda}$ and $H$ can not be unitarily equivalent. In other words, the operator $U_{\lambda}$ can not be the unitary operator. Hence the conclusion that  the operator $T$ defining the operator $U_{\lambda}$ can not be a selfadjoint  operator.

The Mandelstam--Tamm uncertainty relation (\ref{M5}) is also not free of controversies.
People applying and using
the above described derivation of (\ref{M5})
in their discussions of the time-energy uncertainty relation made use (consciously or not) of a hidden assumption that right hand sides of Es. (\ref{M1}), (\ref{M3}) are non--zero, that is that there does not exist any vector $|\phi_{\beta}\rangle \in {\cal H}$ such that $\langle[A,H]\rangle_{\phi_{\beta}}  = 0$,  or $d/dt\langle A\rangle_{\phi_{\beta}} =0$. Although in the original paper of Mandelstam and Tamm \cite{M-T} there is a reservation that for the validity of the formula of the type (\ref{M5}) it is necessary that $\Delta_{\phi} H \neq 0$ (see also, e.g. \cite{Gray,Aharonov}),
there are not an analogous reservations in \cite{M} and in many other papers.

Basic principles  of mathematics require
that before the dividing the both sides of Eq. (\ref{M3})  by $\left|  \frac{d}{dt} \langle A \rangle_{\phi} \right|$, one should check whether $ \frac{d}{dt} \langle A \rangle_{\phi} $ is different from zero or not. Let us do this now: Let $ \Sigma_{H} \subset {\cal H}$  be a set of eigenvectors $ |\phi_{\beta}\rangle $ of $H$ for the eigenvalues $E_{\beta}$. We have   $H|\phi_{\beta}\rangle = E_{\beta}|\phi_{\beta}\rangle$
for all $|\phi_{\beta}\rangle \in \Sigma_{H}$ and therefore for all $|\phi_{\beta}\rangle \in \Sigma_{H}  \bigcap {\cal D}(A)$ (see (\ref{M2})),
\begin{equation}
\langle [A,H] \rangle_{\phi_{\beta}}  = i\hbar  \frac{d}{dt} \langle A \rangle_{\phi_{\beta}} \equiv 0. \label{U1}
\end{equation}
Similarly,
\begin{equation}
\Delta_{\phi_{\beta}} H = \sqrt{\langle H^{2}\rangle_{\phi_{\beta}} - (\langle H\rangle_{\phi_{\beta}})^{2}} \stackrel{\rm def}{=} \Delta_{\phi_{\beta}} E \equiv 0,
\label{U2}
\end{equation}
for all $|\phi_{\beta}\rangle \in \Sigma_{H}$. This means that in all such cases the non--strict inequality (\ref{M3}) takes
the form of the following equality
\begin{equation}
\Delta_{\phi} A \cdot 0 \;= \;\frac{\hbar }{2} \cdot 0. \label{U3}
\end{equation}
In other words, one can not divide the both sides of the inequality (\ref{M3}) by $ \left|\frac{d}{dt} \langle A \rangle_{\phi}\right| \equiv 0 $  for all $|\phi_{\beta}\rangle \in \Sigma_{H}$, because in all such cases the result is  an undefined number and such mathematical operations are unacceptable.
It should be noted that although the authors of the publications \cite{M,Gray} and many others  knew that the property (\ref{U1}) occurs for the vectors from the set $\Sigma_{H}$, it did not prevent them to divide both sides of the inequality (\ref{M3})  by  $\left|  \frac{d}{dt} \langle A \rangle_{\phi} \right|$, that is by  $\left|  \frac{d}{dt} \langle A \rangle_{\phi} \right| \equiv 0$,  without taking into account (\ref{U2}) and without any explanations.
What is more, this shows that there is no reason to think of $\tau_{A}$ as infinity in this case as it was done, e.g, in \cite{M,Gray}.
In general,the problem is that usually the set $\Sigma_{H}$ of the eigenvectors of the Hamiltonian $H$ is a linearly dense (complete) set in the
 state space ${\cal H}$.

Similar picture one meets when $|\phi\rangle = |\phi_{\alpha}\rangle$ is an eigenvector for $A$. (This case was also noticed in \cite{Gray}). Then also for any $|\phi_{\alpha}\rangle \in \Sigma_{A}\bigcap{\cal D}(H)$, (where by $\Sigma_{A}$ we denote the set of eigenvectors $|\phi_{\alpha}\rangle$ for $A$), $\left|  \frac{d}{dt} \langle A \rangle_{\phi} \right| \equiv 0$ and $\Delta_{\phi}A \equiv 0$. Thus, instead of (\ref{U3}) one  once more has,
\begin{equation}
0\cdot \Delta_{\phi}H = \frac{\hbar}{2}\cdot 0, \label{U4}
\end{equation}
and
once again dividing both sides of this inequality  by  zero has no mathematical sense.
Now  note that
the relations (\ref{H1}), (\ref{R1})
are always  satisfied  for   all $|\phi\rangle \in {\cal H}$ fulfilling the conditions specified before Eq. (\ref{R1}).
In contrast to this property, results (\ref{U3}), (\ref{U4}) mean that
we have  proved that the Mandestam--Tamm relation (\ref{M4}) can not be true not only on the set $\Sigma_{H} \subset {\cal H}$, whose  span is usually dense in ${\cal H}$,  but also on the set   $\Sigma_{A} \subset {\cal H}$.

 Hence the conclusion that such relations as (\ref{M4}) and then (\ref{M5}) can not be considered as correct and rigorous seems to be justified.
Summing up, we have proved that
contrary to the uncertainty relations (\ref{H1}) and (\ref{R1}),
the relations of  type (\ref{H2}) and (\ref{M5}) can not hold on  linearly dense sets in
the state space
${\cal H}$  and therefore such relations can not be considered as the universally valid.

\section{Discussion}

Conclusions presented in the previous Section
agrees  with the intuitive understanding of stationary states.  These states are represented by eigenvectors of the Hamiltonian $H$ of the system under considerations and if one knows that the system is in a stationary state represented, assume, by the state vector $|\phi_{\beta}\rangle$ and then $\langle E\rangle_{\phi} \equiv \langle H\rangle_{\phi} = E_{\beta}$ and
$\Delta_{\phi_{\beta}}E = \Delta_{\phi_{\beta}}H =0$ and
then one is sure  that at any time $t$ (and during  any time interval $\Delta t = t_{2} - t_{1} $, where $t_{1} < t_{2} < \infty$) the energy is equal $E_{\beta}$ or that $\Delta E = E_{\beta}(t_{2}) - E_{\beta}(t_{1}) \equiv 0$.

In addition to the doubts discussed above and relating to validity of the time--energy uncertainty relations
a thorough analysis of the relation (\ref{H2}) suggests one more interpretative ambiguity. Namely let us consider the minimal uncertainty version of  (\ref{H2}):
\begin{equation}
\Delta_{\phi} t \cdot \Delta_{\phi} E \;=\; \frac{ \hbar}{2}.\label{H2=}
\end{equation}
Analyzing the ideas expressed in \cite{H,H2}, it can be seen that Heinserberg  was sure that the time-- energy uncertainty relation  is a completely general relation and applies in the quantum world without any exceptions. This means that according to Heisenberg's ideas this relation should be also valid in the case of photons. Then,
let us invoke a much older  relation, namely the Planck--Einstein relation:
\begin{equation}
E_{\phi} = h\nu_{\phi},  \label{hn}
\end{equation}
(where $h$ is the Planck's constant nad $\nu_{\phi}$ is the frequency),
which constituted one of the foundations enabling the emergence of quantum mechanics. This relation  plays still a fundamental role in Quantum Theory and it was verified many times using direct and indirect methods.

 The frequency $\nu_{\phi}$ is connected with the period $\mathcal{T}_{\phi}$ by the relation
\begin{equation}
\nu_{\phi} = \frac{1}{\mathcal{T}_{\phi}}. \label{T}
\end{equation}
Now using the last relation (\ref{T}) one can rewrite the Planck--Einstein relation (\ref{hn}) as follows:
\begin{equation}
\mathcal{T}_{\phi}\,E_{\phi}\;=\;h, \label{TE=h}
\end{equation}
which means that there is,
\begin{equation}
 \mathcal{T}_{\phi}\,E_{\phi}\; > \; \frac{\hbar}{2}. \label{TE>h}
\end{equation}
From the mathematical point of view equations (\ref{H2=}) and (\ref{TE=h}) are identical. (To be more precise:  the equation (\ref{TE=h}) is a re--scaled version of the equation (\ref{H2=}) and scaling factor equals $\frac{1}{4\pi} \simeq 0.08$). On the other hand the inequality (\ref{TE>h}) is the strong case of the Heisenberg inequality (\ref{H2}).
The problem is that relations (\ref{TE=h}) (and (\ref{hn})) combine exact values of time $t= \mathcal{T}_{\phi}$ and energy  $E_{\phi}$ (or $E_{\phi}$ and $\nu_{\phi}$) with each other while the equation (\ref{H2=}) combines uncertainties of time $t$ and energy $E$. In the light of this analysis the standard interpretation of the Heisenberg relation (\ref{H2=}) (and (\ref{H2})) may not be obvious and correct.
Simply:
Equation (\ref{TE=h}) says that if one find that the exact value of the energy of the photon in the state $|\phi\rangle$ is $E_{\phi}$  ,
then one is sure that the period is $\mathcal{T}_{\phi}$ (or that the frequency is $\nu_{\phi} \,=\, 1/\mathcal{T}_{\phi}$) and, of course because  the value of $E$ is exact  then  in this case there must be $\Delta_{\phi} E =0$.
At the same time, equation (\ref{H2=})  and inequality (\ref{H2}) state  that if the value of $E$ is exact and thus $\Delta_{\phi} E =0$ then  simultaneously there must  be $\Delta_{\phi}t = \infty$, which means that it should be impossible to determine the exact value of the period $\mathcal{T}_{\phi}$ or frequency $\nu_{\phi}$.

It was signaled earlier  there is a reservation in \cite{M-T} that derivation of (\ref{M5}) does not go for eigenvectors of $H$ (Then $\Delta H = 0$): it can be only applied
for eigenvectors corresponding to the continuous part of the spectrum of $H$. As an example of possible applications of the relation (\ref{M5}) unstable states modeled by wave--packets  of such eigenvectors of $H$ are considered in \cite{M-T}, where using (\ref{M5}) the  relation connecting half--time $\tau_{1/2}$ of the unstable state, say $|\varphi\rangle$, with the uncertainty $\Delta_{\varphi} H$ was found: $\tau_{1/2}\,\cdot\, \Delta_{\varphi} H \geq \frac{\pi}{4}\,h$. In general, when one considers unstable states such a  relation and the similar one appear naturally \cite{fock,kb,boy} but this is quite another situation then that described by the relations (\ref{H1}), (\ref{R1}). The other example is
a relation between a life--time $\tau_{\varphi}$ of the system in the unstable state, $|\varphi \rangle$,
and the decay width $\Gamma_{\varphi}$: In such  cases we have $\tau_{\varphi} \cdot \Gamma_{\varphi} = \hbar$ but there are not any uncertainties  of the type  $\Delta E$ and $\Delta t$ in this relation (see, e.g., \cite{fock}). Note that in all such  cases the vector $|\varphi\rangle$ representing the unstable state can not be the eigenvector of the Hamiltonian $H$. In general relations described in this paragraph are rather examples of relations analogous to the relation (\ref{TE=h}) and can not be interpreted as an realization  of the Heisenberg's ideas, i.e. as relation connecting uncertainties  of the time and energy.

\section{Conclusions}

The analysis of the discussion of
relations (\ref{H2}) and (\ref{M5})
in previous Sections together with the conclusions presented in \cite{ku-2019}
show that these time--energy uncertainty relations are not well founded and can not be considered as universally valid.
This means that
using these relations as the basis for predictions of the properties and of a behavior of some systems in physics or astrophysics (including cosmology --- see, e.g., \cite{skr,cos}) one should be very careful interpreting and applying results obtained.
In general in some problems the use of the relation (\ref{M5}) may be reasonable (see, e.g. the case of unstable states) but then they should not be interpreted analogously to the relations (\ref{H1}), (\ref{R1}) (i.e., as the relations connecting the uncertainties) but as the relations of the type (\ref{TE=h}) connecting the exact values of quantities corresponding to the time and energy.

\end{document}